\documentclass{emulateapj}
\newcommand\beq{\begin{equation}}
\newcommand\eeq{\end{equation}}

\shorttitle{M81 Globular Cluster System}
\shortauthors{Ma et al.}

\begin{document}
\slugcomment{PASP, in press}
\title{Metal Abundance Properties of M81 Globular
Cluster System}

\author{
Jun Ma\altaffilmark{1}, David Burstein\altaffilmark{2}, Zhou
Fan\altaffilmark{1,3}, Xu Zhou\altaffilmark{1}, Jiansheng
Chen\altaffilmark{1}, Zhaoji Jiang\altaffilmark{1}, Zhenyu
Wu\altaffilmark{1} and Jianghua Wu\altaffilmark{1}}

\altaffiltext{1}{National Astronomical Observatories, Chinese
Academy of Sciences, Beijing, 100012, P. R. China,
majun@vega.bac.pku.edu.cn}

\altaffiltext{2}{Department of Physics
and Astronomy, Box 871504, Arizona State University, Tempe, AZ
85287--1504}

\altaffiltext{3}{Graduate University of Chinese Academy of
Sciences, 19A Yuquan Road, Shijingshan District, Beijing 100049,
China}

%\authoremail{}

\begin{abstract}
This paper is the third in the series of papers on M81 globular
clusters. In this paper, we present spatial and metal abundance
properties of 95 M81 globular clusters, which comprise nearly half
of all the M81 globular cluster system. These globular clusters
are divided into two M81 metallicity groups by a KMM test. Our
results show that, the metal-rich clusters did not demonstrate a
centrally concentrated spatial distribution as ones in M31, and
metal-poor clusters tend to be less spatially concentrated. In
other words, the distribution of the metal-rich clusters in M81 is
not very similar to that of M31. Most of the metal-rich clusters
distribute at projected radii of 4-8 kpc. It is also noted that
the metal-rich clusters distribute within the inner 20 kpc, and
the metal-poor ones do out to radii of $\sim 40$ kpc. Like our
Galaxy and M31, the metallicity distribution of globular clusters
in M81 along galactocentric radius suggests that some dissipation
occurred during the formation of the globular cluster system, i.e.
smooth, pressure-supported collapse models of galaxies are
unlikely to produce such radial distribution of metallicity
presented in this paper. There is not evident correlation between
globular cluster luminosity and metallicity in M81 globular
clusters. The overwhelming conclusion of this paper seems to be
that a more complete and thorough cluster search is needed in M81.

\end{abstract}


\section{Introduction}

An understanding of galaxy formation and evolution is one of the
principal goals of modern astrophysics. Globular clusters are
fossils of the earliest stages of galaxy formation and evolution.
They are bright, easily recognized packages containing a stellar
population with a homogeneous abundance and age. So, their
integrated properties of location, abundance, and kinematics
provide valuable clues to the nature and duration of galaxy
formation \citep{bh00}.

The metallicity distribution of globular clusters is of particular
importance in deepening our knowledge of the dynamical and
chemical evolution of the parent galaxies. For example, the
globular clusters of many elliptical galaxies show multi-modal
metallicity distributions, suggesting that multiple star formation
episodes occurred in these elliptical galaxies in the past
\citep{zepf93,bh00}. Great progress has been made in the past
decade in our understanding of globular cluster systems of
galaxies, especially the discovery that many galaxies possess two
or more distinct subpopulations of globular clusters
\citep[e.g.,][and references therein]{west04}. Based on the data
from $Hubble$ $Space$ $Telescope$ ($HST$) archive,
\citet{Gebhardt99}, \citet{larsen01} and \citet{kw01} presented
that many large galaxies possess two or more subpopulations of
globular clusters that have quite different chemical compositions.
Recently, \citet{peng06} presented the color distributions of
globular cluster systems for 100 early-type galaxies observed in
the Virgo Cluster Survey with the Advanced Camera for Surveys
(ACS) on the $HST$, and found that, on average, galaxies at all
luminosities in their study appear to have bimodal or asymmetric
GC color/metallicity distributions. The presence of
color-bimodality indicates that there have been at least two major
star-forming mechanisms in the histories of galaxies.

\citet{patr99} presented a metallicity distribution of 133
Galactic globular clusters that apparently shows two peaks (i.e.,
two distinct metal-poor and metal-rich globular cluster
populations). A double-Gaussian can best fit these two
subpopulations, the mean metallicity values are $-1.59$ and
$-0.55$ dex, respectively. Using the data for 247 globular
clusters in M31, \citet{bh00} studied the metallicity
distribution, which is asymmetric, implying the possibility of
bimodality. Then they applied a KMM algorithm showing that the
metallicity distribution is really bimodal. \citet{perrett02}
confirmed the conclusions of \citet{bh00}. \citet{ma05} showed
that the intrinsic $B$ and $V$ colors and metallicities of 94 M81
globular clusters are bimodal, with metallicity peaks at $\rm
{[Fe/H]}\approx -1.45$ and $-0.53$, similar to what we find for
the Milky Way and M31 globular clusters.

M81 is one of the nearest Sa/Sb-type spiral outside the Local
Group, very similar to M31, and roughly as massive as the Milky
Way. So, beyond the Local Group, it is a good candidate for
reaching a detailed study of spiral galaxy globular cluster system
for comparison to the Milky Way and M31 system. \citet{bh91}
derived spectroscopic metallicities for eight globular clusters in
M81 and presented the sample mean of $\rm [Fe/H]=-1.46\pm0.31$.
\citet{pbh95} obtained low signal-to-noise spectra of 82
candidates, 25 of which were confirmed as $bona$ $fide$ M81
globular clusters. They derived the mean metallicity to be $\rm
[Fe/H]=-1.48\pm0.19$ both from the weighted mean of the individual
metallicities, and directly from the composite spectrum of the 25
confirmed globular clusters. To maximize the success rate of the
globular cluster candidate list for the ongoing spectroscopic
observations, \citet{pr95} used an extensive database that
included photometric, astrometric, and morphological information
on 3774 objects covering over a $>50~\rm{arcmin}$ diameter field
centered on M81 to reveal 70 globular cluster candidates.

\citet{sbkhp02} presented moderate-resolution spectroscopy for 16
globular cluster candidates from the list in \citet{pr95}, and
confirmed these 16 candidates as $bona$ $fide$ globular clusters.
They also obtained metallicities for 15 of the 16 globular
clusters. From their results, \citet{sbkhp02} concluded that the
M81 globular cluster system is very similar to the Milky Way and
M31 systems, both chemically and kinematically.

With the superior resolution of the $HST$, M81 is close enough for
its clusters to be easily resolved on the basis of image structure
\citep{cft01}. Thus, using the $B$, $V$, and $I$ bands of $HST$
Wide Field Planetary Camera 2 (WFPC2), \citet{cft01} imaged eight
fields covering a total area of $\sim 40~\rm{arcmin}^{2}$, and
detected 114 compact star clusters in M81, 59 of which are
globular clusters. Based on the estimated intrinsic colors,
\citet{cwl04} found that the M81 globular cluster system has an
extended metallicity distribution, which argues the presence of
both metal-rich and metal-poor globular clusters. \citet{ma05}
then confirmed this conclusion.

The outline of the paper is as follows. In \S~2 we provide some
statistical relationships. The summary is
presented in \S~3.

\section{Properties of Globular Clusters in M81}

\subsection{Sample of globular clusters}

In the first paper of our series, \citet{ma05} studied the
distributions of intrinsic $B$ and $V$ colors and metallicities of
95 M81 globular clusters which are from \citet{pbh95},
\citet{cft01} and \citet{sbkhp02}. This cluster sample includes
nearly half of the M81 total globular cluster population. About
the M81 total globular cluster population, \citet{pr95} estimated
it to be $210\pm30$ by the $BVR$ photometric, astrometric, and
image structure study of the M81 field; \citet{cft01} indicated
the total number of globular clusters in M81 to be $211\pm29$
using globular cluster estimates in various annular bins and
correcting for incompleteness. It is difficult to detect and
confirm globular clusters beyond the Local Group before $HST$
appears. In fact, even in the Local Group, it is also not easy to
detect and confirm globular clusters. For example, the total
number of globular clusters in M31 was estimated to be $460\pm70$
by \citet{bh01}, the largest number of globular clusters used to
study the metal abundance properties of the M31 globular clusters
includes 301 clusters collected by \citet{perrett02}, a little
more than half of the total number. Beyond the Local Group, the
globular cluster sample in M81 collected by \citet{ma05} includes
the largest number of globular clusters comparing to the total
globular clusters in the host galaxy.

In the second paper of our series, \citet{ma06} presented the
spectral energy distributions of 42 M81 globular clusters selected
from \citet{ma05} in 13 intermediate-band filters from 4000 to
10000{\AA}, using the CCD images of M81 observed as part of the
Beijing-Arizona-Taiwan-Connecticut (hereafter BATC) multicolor
survey of the sky, and confirmed the conclusions of
\citet{sbkhp02} that, M81 contains clusters as young as a few
Gyrs, which were also observed in both M31 and M33. In this paper,
we will study the spatial and metal abundance properties of the
M81 globular clusters using the sample globular clusters of
\citet{ma05}. Figure 1 is the image of M81 in filter BATC07
(5785{\AA}) of BATC multicolor survey of the sky, the circles
indicate the positions of the sample clusters.

\begin{figure}
%\resizebox{\hsize}{!}{\rotatebox{0}{\includegraphics{f1.eps}}}
\vspace{0.0cm} \caption{The image of M81 in filter BATC07
(5785{\AA}) and the positions of the sample star clusters. The
center of the image is located at ${\rm
RA=01^h33^m50^s{\mbox{}\hspace{-0.13cm}.}58}$
Dec=30$^\circ39^{\prime}08^{\prime\prime}{\mbox{}\hspace{-0.15cm}.4}$
(J2000.0). North is up and east is to the left.}
\label{fig:one}
\end{figure}

As mentioned above, \citet{ma05} studied the distributions of
intrinsic $B$ and $V$ colors and metallicities of these 95 M81
globular clusters, and first found that the abundance distribution
of the globular cluster system is consistent with a bimodal
distribution with peaks at $\rm {[Fe/H]}\approx -1.45$ and $-0.53$
based on the KMM algorithm of \citet{abz94}. It is true that, the
appearance of the histogram can be ambiguous and misleading with
binned data, however, the KMM algorithm of \citet{abz94} is a
robust method of analysis without relying on binning methods
\citep[see][]{perrett02}. KMM mixture modelling works under the
assumption that the sample data are independently drawn from a
parent population that comprises a mixture of $N$ Gaussian
distributions. \citet{ma05} presented that, for M81 globular
clusters, the posteriori probabilities of group membership
returned by the KMM algorithm assigned 74 clusters to the
metal-poor population and 20 to the metal-rich population
distribution\footnote{Since the globular cluster 96 of
\citet{cft01} has very high $(B-V)_0$ ($(B-V)_0=1.778$), and the
metallicity obtained using the color-metallicity correlation is
too rich (0.95 dex), \citet{ma05} do not include it when
performing the KMM test.}.

\subsection{Spatial distribution}

Figure 2 shows the projected spatial distributions of the
metal-poor and metal-rich globular clusters in M81. The distance
modulus for M81 is adopted to be 27.8 \citep{fwm94,cft01}. We
adopted the inclination and position angles to be $59^{\circ}$ and
$157^{\circ}$ of M81 as \citet{cft01} did, respectively. When the
line of intersection (i.e. the major axis of the image) between
the galactic plane and tangent plane is taken as the polar axis,
it is easily proved that:

\begin{equation}
r=\rho\sqrt{1+\tan^2 \gamma\sin^2\theta}
\end{equation}
and
\begin{equation}
\tan\phi=\frac{\tan\theta}{\cos\gamma},
\end{equation}
where $r$ and $\phi$ are the polar co-ordinates in the galactic
plane, and $\rho$ and  $\theta$ are the corresponding co-ordinates
in the tangent plane, and $\gamma$ is the inclination angle of the
galactic disk. Using formula (1), we can obtain the distances of
our sample clusters from the center of M81, which are listed in
Table 1. In Table 1, we also listed the metallicities of the
sample clusters from \citet{ma05}. From Figure 2, it is clear that
the metal-rich globular clusters in M81 are not as centrally
concentrated as the metal-rich globular clusters of M31 are
\citep{hbk91,perrett02}. Figure 3 presents the histogram for the
metal-poor and metal-rich globular clusters in M81. It shows that
most of the metal-rich clusters distribute at projected radii of
4-8 kpc. It is also noted that the metal-rich clusters distribute
within the inner 20 kpc, and the metal-poor ones do out to radii
of $\sim 40$ kpc. In the Milky Way, the metal-rich GCs reveal
significant rotations and have historically been associated with
the thick-disk system \citep{zinn85,taft89}; however, other works
\citep{fw82,minniti95,patr99,forbes01} suggested that metal-rich
GCs within $\sim 5$ kpc of the Milky Way Galactic center are
better associated with the bulge and bar. In M31, \citet{ew88}
showed that the metal-rich clusters constitute a more highly
flatted system than the metal-poor ones, and appear to have
disklike kinematics; \citet{hbk91} showed that the metal-rich GCs
are preferentially close to the galaxy center. At the same time,
\citet{hbk91} showed that the distinction between the rotation of
the metal-rich and metal-poor clusters is most apparent in the
inner 2 kpc. So, \citet{hbk91} concluded that the rich-metal
clusters in M31 appear to form a central rotating disk system.
With the largest sample of 321 velocities, \citet{perrett02}
provided a more comprehensive investigation on the kinematics of
the M31 cluster system. \citet{perrett02} showed that, the
metal-rich globular clusters of M31 appear to constitute a
distinct kinematic subsystem that demonstrates a centrally spatial
distribution with a high rotation amplitude, but does not appear
significantly flattened, which is consistent with a bulge
population. It is of interests to mention that, \citet{sbkhp02}
performed a maximum-likelihood kinematic analysis on 166 M31
clusters of \citet{bh00} and found that the most significant
difference between the rotation of the metal-rich and metal-poor
clusters occurs at intermediate projected galactocentric radii.
Especially, \citet{sbkhp02} presented a potential thick-disk
population among M31's metal-rich globular clusters. For M81
globular clusters, \citet{sbkhp02} performed a kinematic analysis
of the velocities of 44 M81 globular clusters, and strongly
suggested that the metal-rich clusters are rotating in the same
sense as the gas in the disk of M81. \citet{sbkhp02} concluded
that, although their cluster sample is not large enough to make a
direct comparison between metal-rich and metal-poor clusters in
specific radius ranges, the conclusion with M81's metal-rich
clusters at intermediate projected radii being associated with a
thick disk in M81 is correct. It is true that, from Figure 3, most
of metal-rich clusters distribute at projected radii of 4-8 kpc.
So, at least, we can conclude that most of the metal-rich clusters
in our sample are not associated with a bulge cluster system of
M81; they may associate with a thick disk in M81 as indicated by
\citet{sbkhp02}. The sample clusters of this paper include all the
sample clusters of \citet{sbkhp02}. But, except for the sample
clusters of \citet{sbkhp02}, the other clusters have no published
radial velocity estimates. So, obviously, more kinematic and
metallicity data are needed for globular clusters in M81 to
determine if the inner metal-rich GCs have kinematic properties
that are consistent with the bulge and the metal-rich GCs at
projected radii of 4-8 kpc are associated with a thick disk in
M81.

\begin{figure}
\resizebox{\hsize}{!}{\rotatebox{-90}{\includegraphics{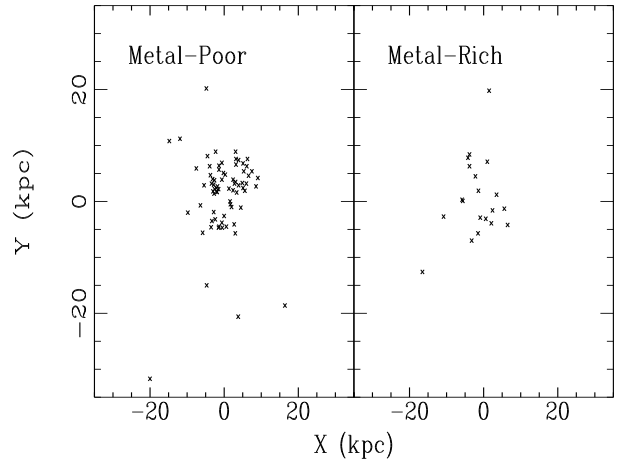}}}
\vspace{0.0cm} \caption{Spatial distributions of the metal-rich
and metal-poor globular clusters.} \label{fig:two}
\end{figure}

\begin{figure}
\resizebox{\hsize}{!}{\rotatebox{-90}{\includegraphics{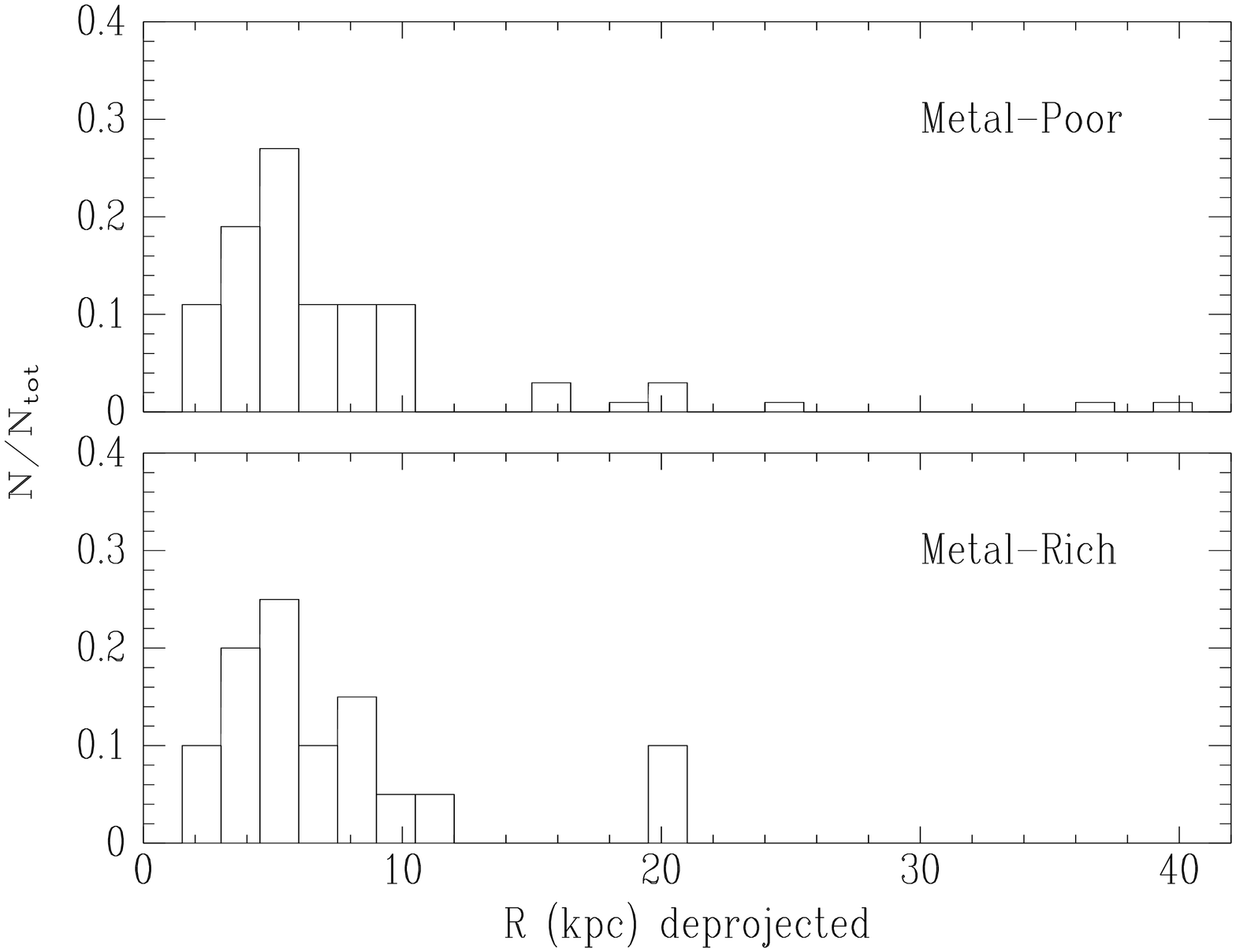}}}
\vspace{0.0cm} \caption{Radial distributions of the metal-rich and
metal-poor globular clusters.} \label{fig:three}
\end{figure}

\subsection{Metallicity gradient}

\begin{figure}
\resizebox{\hsize}{!}{\rotatebox{-90}{\includegraphics{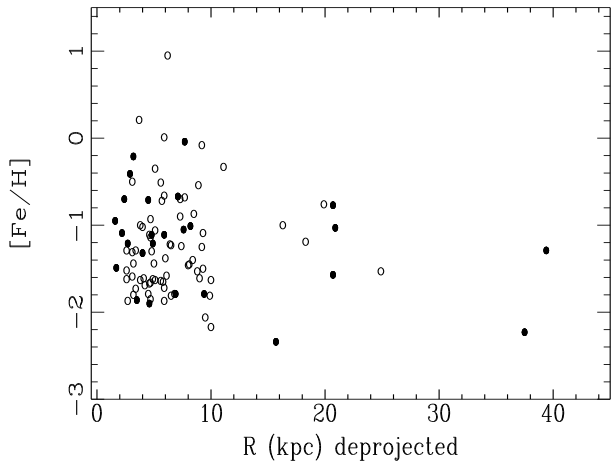}}}
\vspace{0.0cm} \caption{Metallicity as function of projected
radius for M81 globular clusters. Black circles indicate the
clusters with spectroscopic metallicities with uncertainties
smaller than 1.0 dex.} \label{fig:four}
\end{figure}

The presence or absence of a radial trend in globular cluster
metallicities is an important test of galaxy formation theory
\citep{bh00}. If a galaxy forms as a consequence of a monolithic
dissipative and rapid collapse of a single massive,
nearly-spherical spinning gas cloud in which the enrichment
timescale is shorter than the collapse time, the halo stars and
globular clusters should show large-scale metallicity gradients
\citep{eggen62,bh00}; however, \citet{sz78} presented a chaotic
scheme in the early evolution of a galaxy, in which loosely bound
pre-enriched fragments merge with the main body of the
proto-galaxy over a significant period, so there should be
homogeneous metallicity distribution. For the Milky Way,
\citet{taft89} showed some evidence that metallicity gradients
with both distance from the Galactic plane and distance from the
Galactic center were present in the disk cluster system. For M31,
there are some inconsistent conclusions, such as \citet{van69}
showed that there is little or no evidence for a correlation
between metallicity and projected radius, but most of his clusters
were inside $50{\arcmin}$; however, some authors \citep[see
e.g.][]{hsv82,sha88,hbk91,perrett02} presented that there is
evidence for a weak but measurable metallicity gradient as a
function of projected radius. \citet{bh00} confirmed the latter
result based on their large sample of spectral metallicity and
color-derived metallicity. Figure 4 plots the metallicity of the
M81 globular clusters as a function of galactocentric radius, in
which black circles indicate the clusters with spectroscopic
metallicities with uncertainties smaller than 1.0 dex. Clearly,
the dominant feature of this diagram is the scatter in metallity
at any radius. At the same time, our sample clusters are mainly
distributed in the inner 10 kpc. So, it is difficult to determine
the metallicity gradient. It is true that, smooth,
pressure-supported collapse models of galaxies are unlikely to
produce a result like this. However, in order to present a
quantitative conclusion, we made least-squares fits: the total
sample of globular clusters does not have a significant
metallicity gradient ($-0.009\pm0.009$ dex $\rm kpc^{-1}$), and
the clusters with spectroscopic metallicities with uncertainties
smaller than 1.0 dex have a marginally significant gradient
($-0.018\pm0.01$ dex $\rm kpc^{-1}$). This result is in agreement
with \citet{kong00}, who obtained metallicity maps of M81 field by
comparing simple stellar population synthesis models of BC96
\citep{bc96} with the integrated photometric measurements of the
BATC photometric system, and did not find, within their errors,
any obvious metallicity gradient from the central region to the
bulge and disk of M81. But, we should emphasize that, in the
least-squares fits of this paper, the metal-rich clusters seem to
act an important part in determining the metallicity gradient.
From Figure 4, we can also see the decrease in the ``upper
envelope" of metallicity reported by \citet{hbk91} and
\citet{bh00} for M31 globular clusters. In fact, we do a
least-squares fits for the metal-poor clusters of M81, the
metallicity gradient is $-0.006\pm0.006$ dex $\rm kpc^{-1}$. This
metallicity gradient may not have any statistical meanings. It is
clear that, the small sample of rich-metal clusters in M81 cannot
present any firm conclusion about metallicity gradient.

\subsection{Metallicity versus absolute magnitude}

\begin{figure}
\resizebox{\hsize}{!}{\rotatebox{-90}{\includegraphics{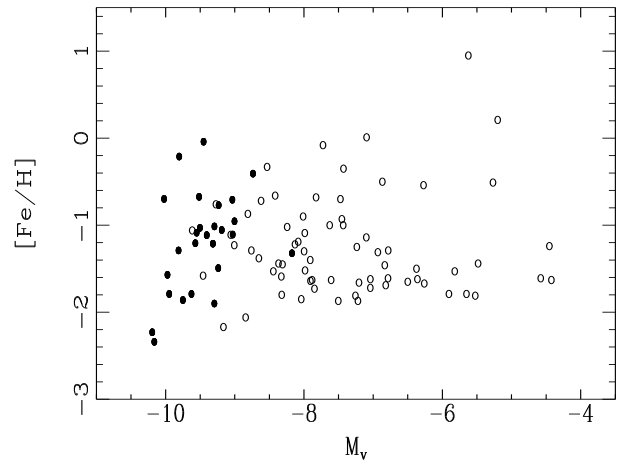}}}
\vspace{0.0cm} \caption{Metallicity versus absolute magnitude for
M81 globular clusters. Black circles indicate the clusters with
spectroscopic metallicities with uncertainties smaller than 1.0
dex.} \label{fig:five}
\end{figure}

The correlation between cluster mass (or luminosity) and
metallicity is important in globular cluster formation theory.
%As \citet{bh00} pointed out that,
It is generally believed that, if self-enrichment is important in
globular clusters, the most massive clusters could retain their
metal-enriched supernova ejecta, so the metal abundance should
increase with cluster mass; the opposite is true if cooling from
metals determines the temperature in the cluster-forming clouds
\citep{bh00}. The self-enrichment of GCs has been in detail
studied in some aspects \citep[see detail from][]{sbsb06}.
However, it is interesting of mentioning the model of globular
cluster self-enrichment developed by \citet{pj99}. In this model,
cold and dense clouds embedded in the hot protogalactic medium are
assumed to be the progenitors of galactic halo globular clusters.
Based on this model, \citet{pg01} presented that the most
metal-rich proto-globular clusters are the least massive ones.

For M31 globular clusters, \citet{hbk91} first presented the
metallicity versus apparent magnitude for 150 M31 clusters, and
did not find any trend of metallicity with luminosity; then,
\citet{bh00} showed metallicity versus dereddened apparent
magnitude using a large cluster sample including 247 objects, and
confirmed the conclusion of \citet{hbk91}.

Figure 5 shows metallicity versus absolute magnitude for M81
globular clusters, in which black circles indicate the clusters
with spectroscopic metallicities with uncertainties smaller than
1.0 dex. The metallicity and absolute magnitudes for the sample
clusters are from \citet{ma05}. It is true that there is not
obvious trend of metallicity with luminosity as M31 GCs do.
Least-squares fits show no evidence for a relationship between
luminosity and metallicity in M81 clusters.

As we know, $HST$ provides a unique tool for studying globular
clusters in extragalaxies. Recently, based on the ACS on the
$HST$, \citet{harris06} and \citet{sbsb06} found that, in giant
ellipticals such as M87, NGC4649 and NGC7094, luminous blue GCs
(i.e. metal-poor GCs) reveal a trend of having redder colors, such
that more massive GCs are more red (metal-rich). This trend is
referred to as a `blue tilt' \citep[also see][]{strader06}. This
`blue tilt' was interpreted as a result of self-enrichment
\citep{sbsb06}. \citet{sbsb06} speculatively suggested that these
GCs once possessed dark matter halos. \citet{spitler06}
subsequently found that this `blue tilt' is also true in the
Sombrero spiral galaxy (NGC4594) and may extend to less luminous
GCs with a somewhat shallower slope that was derived by
\citet{harris06} and \citet{sbsb06}. As \citet{spitler06} pointed
out that, the Sombrero provides the first example of this trend in
a spiral galaxy and in a galaxy found in a low-density galaxy
environment. However, in these ACS studies, the metal-rich (red)
GCs did not show this corresponding trend \citep[also
see][]{bekki07}. Based on high-resolution cosmological simulation
with globular clusters, \citet{bekki07} investigated formation
processes and physical properties of globular cluster system in
galaxies, and found that, luminous metal-poor clusters would
emerge a correlation between luminosity and metallicity if they
originated from nuclei of low-mass galaxies at high $z$. In fact,
in the simulations of \citet{bekki07}, the `simulated blue tilts'
emerge from the assumption that luminous metal-poor clusters
originate from stellar galactic nuclei of the more massive
nucleated galaxies with a luminosity-metallicity relation. So, it
is evident that, in \citet{bekki07}, galaxies which experienced
more accretion/merging events of nucleated low-mass galaxies are
more likely to show a blue tilt \citep[see details
from][]{bekki07}.

\section{Summary}

In this paper we present spatial and metal abundance properties of
95 M81 globular clusters, which are collected by \citet{ma05}.
This cluster sample includes nearly half of the M81 total globular
cluster population, is the largest one comparing to the total
globular clusters of the host galaxy beyond the Local Group. Our
conclusions are as follows:

1. The metal-rich clusters did not demonstrate a centrally
concentrated spatial distribution as ones in M31, and metal-poor
clusters tend to be less spatially concentrated. Most of
metal-rich clusters distribute at projected radii of 4-8 kpc. We
can conclude that most of the metal-rich clusters in our sample
are not associated with a bulge cluster system of M81; they may
associate with a thick disk in M81 as indicated by
\citet{sbkhp02}.

2. The globular clusters in M81 have a small radial metallicity
gradient like M31 and our Galaxy, suggesting that some dissipation
occurred during the formation of the globular cluster system.

3. There is not obvious trend of metallicity with luminosity in
M81 globular clusters.

\acknowledgments We are indebted to the referee for his/her
thoughtful comments and insightful suggestions that improved this
paper greatly. This work has been supported by the Chinese
National Natural Science Foundation Nos 10473012, 10573020,
10633020, 10673012 and 10603006; and by National Basic Research
Program of China (973 Program) No. 2007CB815403.

\clearpage
\setcounter{table}{0}
\begin{table*}
\caption{Globular cluster sample and properties}
%\vspace {0.1cm}
\begin{tabular}{lcc|lcc}
\hline
\hline
${\rm ID^{a}}$ & [Fe/H] & Distance  &${\rm ID^{a}}$ & [Fe/H] & Distance  \\
     &           & from M81 center &      &     & from M81 center\\
     &           &    (kpc)       &       &     &  (kpc)  \\
\hline
   Id30244 &    $-1.53\pm 0.072$ &   24.9 &   CFT5 & $-0.87\pm0.070$ &   8.5 \\
   Is40083 &    $-1.29\pm 0.80 $ &   39.4 &   CFT6 & $-1.40\pm0.084$ &   8.4 \\
   Is40165 &    $-1.57\pm 0.43$  &   20.7 &   CFT8 & $-0.90\pm0.060$ &   7.3 \\
   Is40181 &    $-0.76\pm 0.072$ &   19.9 &  CFT15 & $-1.81\pm0.314$ &   9.9 \\
   Is50037 &    $-2.34\pm 0.83$  &   15.7 &  CFT16 & $-1.53\pm0.360$ &   8.8 \\
   Is50225 &    $-0.04\pm 0.59$  &    7.7 &  CFT20 & $-1.46\pm0.205$ &   8.0 \\
   Is50233 &    $-1.23\pm 0.072$ &    6.5 &  CFT21 & $-1.50\pm0.263$ &   9.3 \\
   Is50286 &    $-1.45\pm 0.072$ &    8.1 &  CFT22 & $-0.70\pm0.253$ &   7.3 \\
   Id50357 &    $-0.33\pm 0.072$ &   11.1 &  CFT28 & $-1.24\pm0.393$ &   7.4 \\
   Is50394 &    $-2.17\pm 0.072$ &   10.0 &  CFT30 & $-1.09\pm0.128$ &   9.3 \\
   Id50401 &    $-0.72\pm 0.072$ &    5.7 &  CFT31 & $-0.54\pm0.502$ &   8.9 \\
   Id50415 &    $-1.90\pm 0.71$  &    4.6 &  CFT32 & $-0.08\pm0.205$ &   9.2 \\
   Id50696 &    $-1.86\pm 0.50$  &    3.5 &  CFT34 & $-1.25\pm0.174$ &   9.2 \\
   Id50785 &    $-1.58\pm 0.072$ &    6.1 &  CFT37 & $-1.80\pm0.031$ &   3.2 \\
   Is50861 &    $-1.38\pm 0.072$ &    6.0 &  CFT38 & $-1.44\pm0.029$ &   3.2 \\
   Is50886 &    $-1.79\pm 0.87$  &    6.9 &  CFT39 & $-1.29\pm0.022$ &   3.4 \\
   Id50960 &    $-1.79\pm 0.64$  &    9.4 &  CFT41 & $-1.59\pm0.022$ &   3.1 \\
   Is51027 &    $-2.06\pm 0.072$ &    9.5 &  CFT42 & $-1.52\pm0.034$ &   2.6 \\
   Is60045 &    $-1.03\pm 0.97$  &   20.9 &  CFT43 & $-1.62\pm0.138$ &   2.6 \\
   Id70319 &    $-1.00\pm 0.072$ &   16.3 &  CFT44 & $-1.61\pm0.123$ &   4.1 \\
   Id70349 &    $-1.19\pm 0.072$ &   18.3 &  CFT45 & $-1.63\pm0.043$ &   5.1 \\
   Is80172 &    $-0.77\pm 0.68$  &   20.7 &  CFT46 & $-1.66\pm0.051$ &   4.7 \\
   Is90103 &    $-2.23\pm 0.99$  &   37.5 &  CFT49 & $-1.67\pm0.191$ &   4.6 \\
    SBKHP1 &    $-1.207\pm 0.369$ &   4.9 &  CFT51 & $-0.35\pm0.084$ &   5.1 \\
    SBKHP2 &    $-0.707\pm 0.167$ &   4.5 &  CFT53 & $-1.81\pm0.058$ &   6.5 \\
    SBKHP3 &    $-0.211\pm 0.193$ &   3.2 &  CFT56 & $-1.63\pm0.034$ &   3.8 \\
    SBKHP4 &    $-0.407\pm 0.088$ &    2.9 &  CFT58 &  $0.21\pm0.343$ &   3.7 \\
    SBKHP5 &    $-1.086\pm 0.091$ &    2.2 &  CFT62 & $-1.11\pm0.017$ &   4.6 \\
    SBKHP6 &    $-1.493\pm 0.206$ &    1.7 &  CFT63 & $-1.69\pm0.628$ &   4.2 \\
    SBKHP7 &    $-0.955\pm 0.098$ &    1.6 &  CFT65 & $-1.87\pm0.080$ &   2.7 \\
    SBKHP8 &    $-0.698\pm 0.058$ &    2.4 &  CFT66 & $-1.85\pm0.077$ &   4.7 \\
    SBKHP9 &    $-1.212\pm 0.133$ &    2.7 &  CFT67 & $-1.31\pm0.092$ &   3.1 \\
   SBKHP10 &    $-1.322\pm 0.356$ &    4.0 &  CFT68 & $-1.62\pm0.060$ &   4.9 \\
   SBKHP11 &    $-1.114\pm 0.409$ &    4.8 &  CFT74 & $-1.72\pm0.060$ &   5.9 \\
   SBKHP12 &    $-1.06\pm 0.072$  &    5.1 &  CFT75 & $-1.64\pm0.046$ &   5.6 \\
   SBKHP13 &    $-1.055\pm 0.062$ &    7.6 &  CFT76 & $-1.87\pm0.046$ &   5.9 \\
   SBKHP14 &    $-1.107\pm 0.074$ &    5.9 &  CFT80 & $-1.79\pm0.152$ &   6.8 \\
   SBKHP15 &    $-1.014\pm 0.713$ &    8.2 &  CFT83 & $-1.63\pm0.282$ &  10.0 \\
   SBKHP16 &    $-0.674\pm 0.044$ &    7.1 &  CFT85 & $-1.61\pm0.203$ &   9.0 \\
     CFT87 &    $-0.66\pm0.087$ &    5.9 &  FT106 & $-0.93\pm0.147$ &   4.7 \\
     CFT90 &    $-0.51\pm1.733$ &    5.6 &  FT108 & $-1.00\pm0.123$ &   3.8 \\
     CFT96 &     $0.95\pm0.792$ &    6.2 &  FT109 & $-0.50\pm0.234$ &   3.1 \\
     CFT97 &    $-1.22\pm0.056$ &    6.4 &  FT110 & $-1.29\pm0.215$ &   2.6 \\
     FT101 &    $-1.44\pm0.386$ &    5.0 &  FT111 & $-1.14\pm0.150$ &   4.7 \\
     FT102 &     $0.01\pm0.659$ &    5.9 &  FT112 & $-1.79\pm0.256$ &   4.5 \\
     FT103 &    $-1.65\pm0.263$ &    5.8 &  FT113 & $-0.68\pm0.140$ &   7.7 \\
     FT104 &    $-1.30\pm0.082$ &    4.8 &  FT114 & $-1.73\pm0.159$ &   3.4 \\
     FT105 &    $-1.02\pm0.075$ &    4.0 &        &       &       \\
\hline
\end{tabular}
\end{table*}

\end{document}